# Fast undersampled dynamic MRI reconstruction using explicit representation learning with Gaussian splatting

M.L. Terpstra[1] and C.A.T. van den Berg[1]
[1]Computational Imaging Group for MRI Therapy & Diagnostics, Center of Image Sciences, University Medical Center Utrecht, the Netherlands

## Synopsis

**Motivation**: Quickly obtaining high-quality MRI from accelerated acquisitions is important to mitigate motion artifacts, maintain patient comfort, and improve clinical efficiency.
**Goals**: To obtain high-quality dynamic MRI using efficient, personalized models.
**Approach**: We propose a novel explicit representation learning approach using Gaussian splatting. Multiple Gaussian primitives are trained to represent the underlying tissue. We extend the Gaussian splatting framework to model anatomical motion, enabling learning an efficient, explicit representation of dynamic MRI.
**Results**: Gaussian splats can be trained in 60s with 0.5ms/dynamic inference time. High-quality cardiac MRI is obtained at R=16. We show that the properties of the Gaussians directly encode physiological properties.

## Impact

Quickly obtaining high-quality MRI from accelerated acquisitions is important to mitigate motion artifacts, maintain patient comfort, and improve clinical efficiency. We show that explicit representation learning using Gaussian splatting enables high-quality cardiac MRI, while directly encoding the underlying tissue properties.

## Introduction

Recently, implicit neural representation (INR) learning has been proposed to overcome the need for large, high-quality datasets required by traditional, supervised deep learning approaches[1-3]. By learning scan-specific continuous functions that map input coordinates to the corresponding Fourier components, high-quality MRI is obtained. However, these continuous functions leverage minimal prior information, resulting in long training times (15-120 minutes)[1-3]. Moreover, because the learned representations are embedded in the nonlinear weights of a multilayer perceptron, they can only be sampled using serial input coordinate enumeration, making inference time-consuming (~30s[1]) while making the extraction of novel properties from these representations extremely challenging. Instead, we propose learning an explicit spatiotemporal representation of the underlying anatomy using Gaussian splatting[4,5]. With this explicit representation, we impose a strong inductive bias, enable fast inference, while accurately capturing spatio-temporal, nonlinear tissue properties.

## Methods

Instead of representing a 2D dynamic MRI acquisition as T×N pixels, 2D Gaussian splatting builds an image representation using M Gaussian primitives (M≪TN)[5]. The intensity propagated by a Gaussian splat G to a pixel p is $G(p) = c \cdot \exp\left(-\frac{1}{2}(\mu - p)^T \Sigma^{-1}(\mu - p)\right)$, where $c \in \mathbb{C}$ is the complex-valued intensity, $\mu \in \mathbb{R}^2$ is the center location, and $\Sigma \in \mathbb{R}^{2\times 2}$ is the covariance matrix, which is factorized using a Cholesky decomposition as $\Sigma = LL^T$ to ensure $\Sigma$ is a symmetric positive-definite matrix[5]. The final image is formed by rasterizing M Gaussian splats, where the value at pixel p is $C_p = \sum_{g \in G} G_g(p)$. Note that, unlike INRs, this rasterization occurs in parallel, enabling fast inference. When subjected to physiological motion, tissue can translate or deform. To represent these motion-induced tissue changes, we propose two extensions to the 2D Gaussian splatting (2DGS) framework[5]:

- We represent Lagrangian motion fields as $\mu_t = \mu + \delta\mu_t$, with $\delta\mu \in \mathbb{R}^{T\times 2}$, enabling rigid translation of the Gaussian splats to accommodate non-rigid translation of tissue,

- We represent $\Sigma_t = (L + \ell_t)(L + \ell_t)^T$, with $\ell \in \mathbb{R}^{T\times 3}$, enabling deformation of the underlying tissue by allowing the shape of the Gaussian splat to vary over time, potentially encoding the underlying tissue properties.

An example of motion-resolved Gaussian splatting is shown in **Figure 1.**

We trained the proposed Gaussian splatting framework for undersampled 2D cardiac MRI reconstruction on the OCMR dataset[6]. We retrospectively undersampled the fully-sampled k-space using the GRO undersampling pattern[7] up to R=32. Coil sensitivity maps (CSMs) were computed using ESPIRiT[8]. The model was trained to minimize $L = ||s_t - M_t FCG_t||^2_{HDR} + \lambda_\mu ||\delta\mu_t||_* + \lambda_\mu ||TV_t(\delta\mu)||_1 + \lambda_c ||TV_t(\ell)||_1$ where $s_t$ is the measured k-space at timepoint t, $M_t$ is the corresponding GRO sampling mask, F is the Fourier operator, C is the CSM operator, $G_t$ is the rasterization of the whole image using all Gaussian splats at timepoint t, $||\cdot||^2_{HDR}$ is the HDR-loss[1], $||\cdot||_*$ is the nuclear norm, $TV_t$ is the total variation in time operator, $||\cdot||_1$ is the L1 norm, and $\lambda_\mu=\lambda_c=1e^{-4}$ are the regularization weights. The number of Gaussian splats used to represent ≈390×150=58500 pixels of a single dynamic varied between 1000, 2000, 3000, 5000, 10000, and 25000 primitives. An example of motion-resolved Gaussian splatting is shown in **Figure 1**. This model was stochastically trained for 7500 steps using the Adam optimizer (lr=1e$^{-3}$). Every 500 steps, Gaussians encoding a small area or low magnitude were removed. The reconstruction quality was measured by computing the SSIM, LPIPS[9], and SNR metrics relative to the fully-sampled reference. Moreover, we compared the reconstructions to a baseline L+S algorithm[10] ($\lambda_L=\lambda_S=0.01$). Finally, we evaluated tissue properties learned by the Gaussian splats by investigating the ratio of eigenvalues of $\Sigma_t$, and the super-resolution potential by rasterizing the Gaussian splats on a denser grid.

## Results

The 2DGS approach was trained in approximately 60 seconds per dataset. Rasterizing all Gaussians takes approximately 0.5 ms per dynamic. Example reconstructions are shown in **Figure 2**, showing that high-quality dynamic MRI at R=16 is feasible. Quantitative evaluation shows that high image quality is maintained at R=16 (SSIM 0.935±0.044, LPIPS 0.162±0.150), outperforming the L+S baseline (**Figure 3**). Moreover, the denoising properties of Gaussian splatting result in higher SNR from ~98.1dB to ~284dB. Analyzing the covariance matrix of the Gaussians reveals properties strongly associated with flow and myocardial contractions (**Figure 4**). Gaussian splatting is amenable to super-resolution, maintaining high-quality imaging and smoother structures than pixel-based upsampling for a 20x denser grid (**Figure 5**).

## Discussion & Conclusion

We have presented a novel framework for dynamic MRI reconstruction using 2D Gaussian splatting. With this framework, highly accelerated MRI can be quickly reconstructed (60s of training, <1ms inference time). At R=16, high image quality is achieved while maintaining accurate cardiac dynamics. Future work will investigate 3D Gaussian splatting, further investigate the tissue properties encoding, and explore the Gaussian splatting structure to improve super-resolution MRI.

# Figures

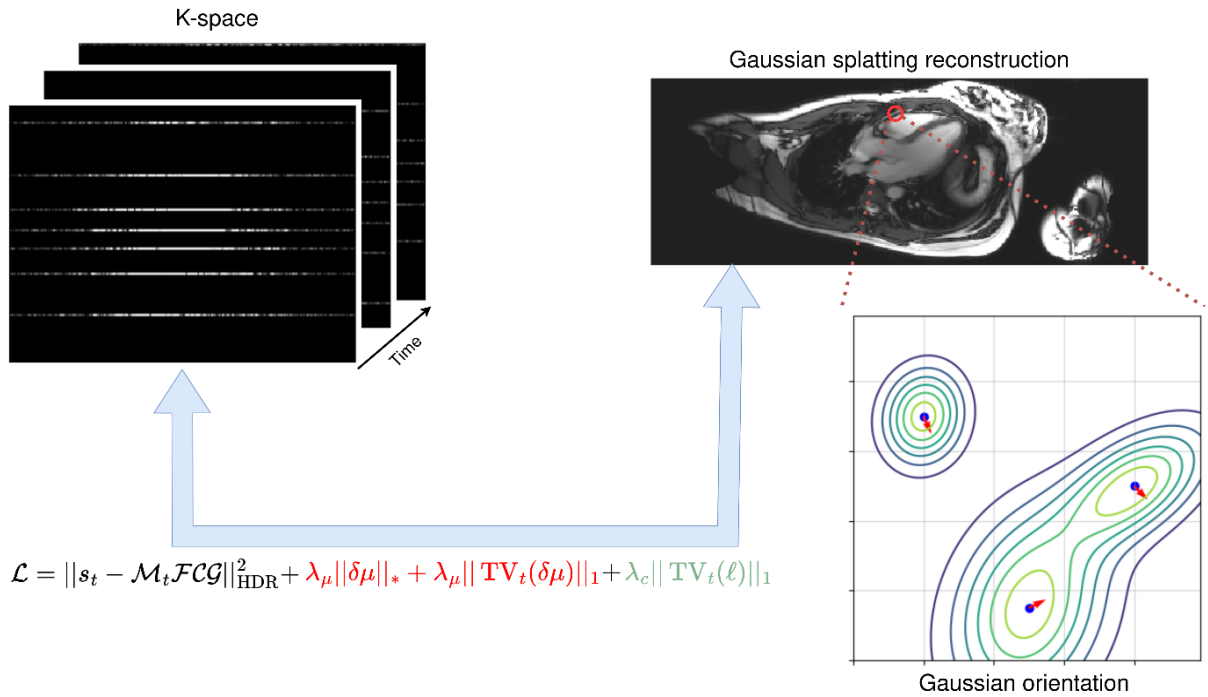

**Figure 1: Fitting and representation.** Rather than representing the image as discrete pixels, the Gaussians (bottom) form an explicit learned representation of the underlying tissue. The centers (blue) are continuous in space, while the covariance matrix determines their shape and size. Each Gaussian is associated with a motion vector, enabling rigid transformation. During rasterization, the Gaussian is splatted to pixel space (grid). Minimizing the loss fits the Gaussians to the dynamic k-space

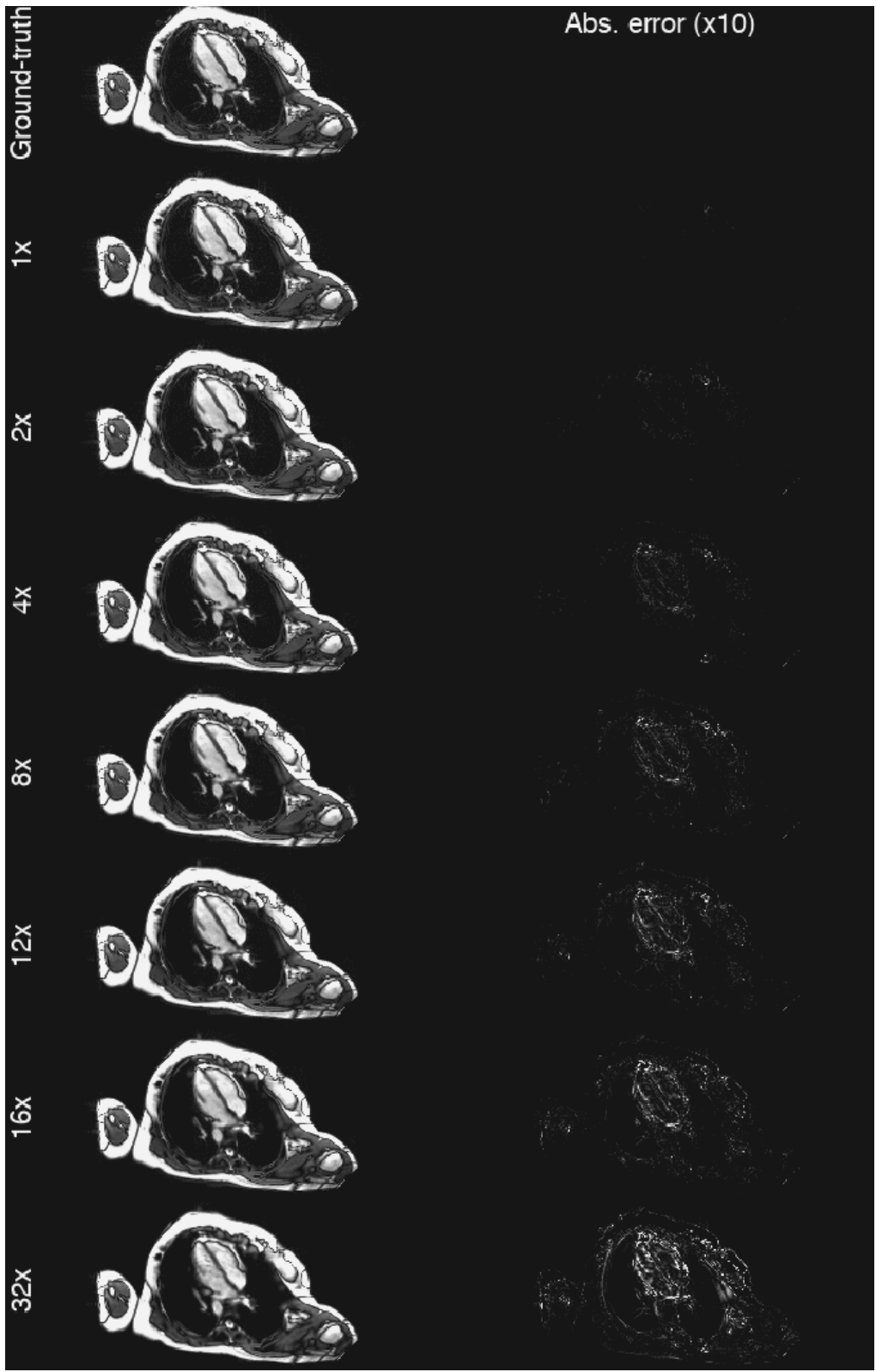

**Figure 2: Reconstruction examples.** The left column: time-series reconstructions. Right column: squared absolute error (x10) versus fully-sampled reference. Top-row is the fully-sampled reference, subsequent rows show Gaussian splatting reconstructions at R=1-32. High reconstruction quality is maintained for the quasi-static regions, while representative cardiac motion is maintained until R=16. However, reduced motion magnitude and image artifacts can be observed at R=32. The animated version of this figure can be found here: https://surfdrive.surf.nl/s/eiRzeNtJSmrd2XJ

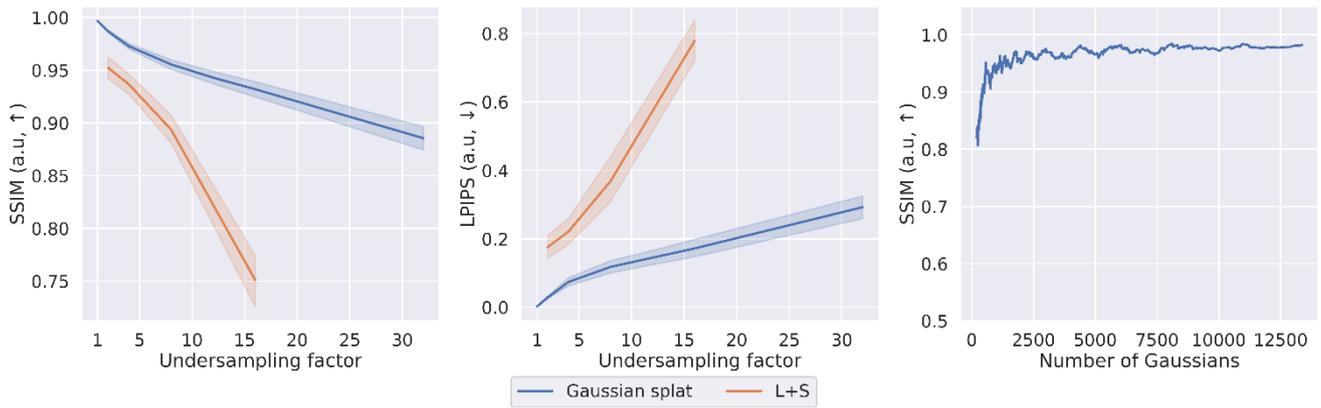

**Figure 3: Quantitative evaluation.** Gaussian splatting outperforms L+S and maintains a high performance, even at high undersampling factors as indicated by the high SSIM score (left) and low LPIPS score (middle). The reconstruction quality is only marginally increases when ≥6000 splats are used to represent the cine data after removing small or low-magnitude Gaussians (right), which is significantly less than average number of pixels in a single dynamic (390×150=58500)

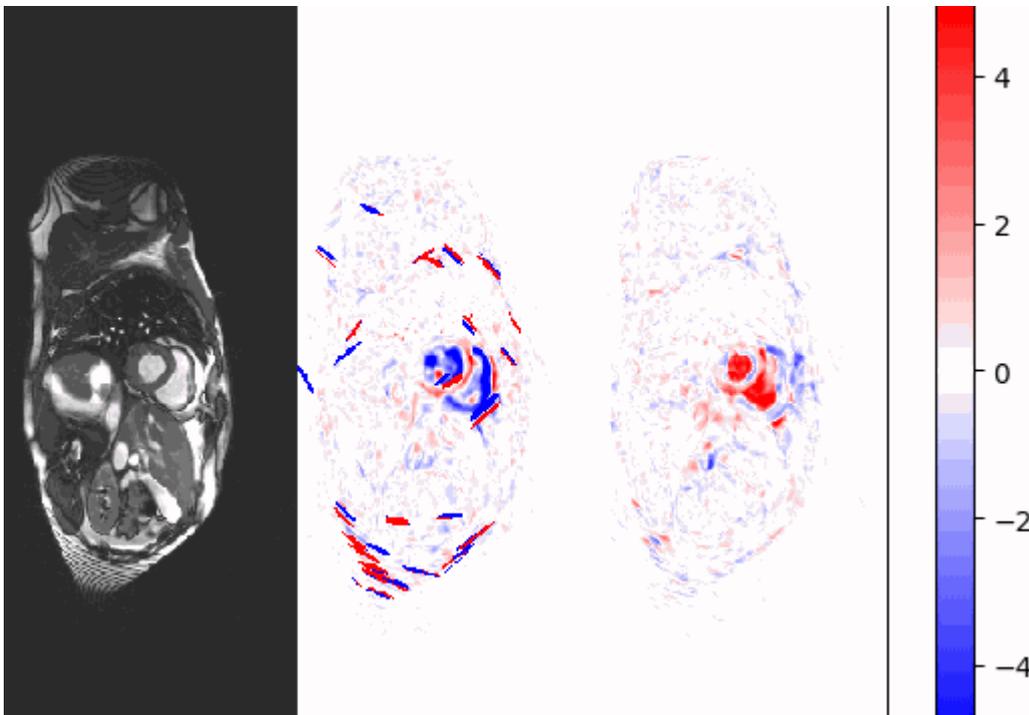

**Figure 4: Encoding tissue parameters**. On the left, the ground-truth reconstruction. In the middle, we splat the ratio of the eigenvalues of relative to the first frame, showing strong representation of myocardial compression. On the right, the determinant of Σ relative to the first frame, showing strong encoding of blood flow through the atrium and ventricle. This shows that Gaussian splatting can potentially learn the underlying tissue properties. The animated version of this figure can be found here: https://surfdrive.surf.nl/s/wPGnFBQXtHxmRrD

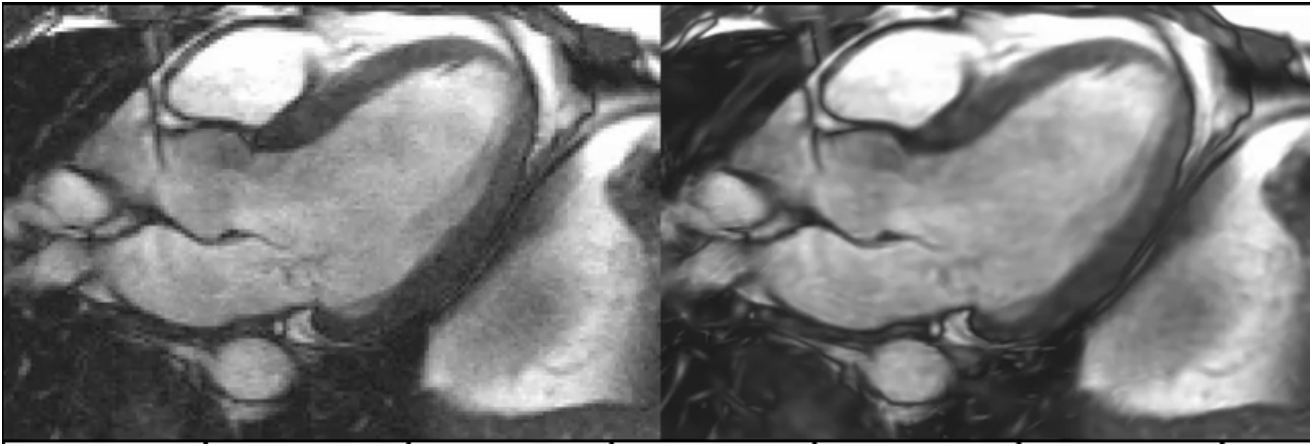

**Figure 5: Figure 5: Super resolution MRI.** On the left, traditional image-based upsampling. On the right, evaluation of the Gaussian splats on 20x denser grid. Notice the smoother edges of the atria and arteries and smooth separation between the blood pool and the myocardium as there is no staircase upsampling effect. The animated version of this figure can be found here: https://surfdrive.surf.nl/s/KtKKmdQKXMMfnTW